\documentstyle[aps,preprint]{revtex}
\begin{document}
\title{Mesoscopic Effects in the Fractional Quantum Hall Regime:
Chiral Luttinger Liquid versus Fermi Liquid} 
\author{Michael R. Geller, Daniel Loss, and George Kirczenow}
\address{Department of Physics, Simon Fraser University, Burnaby B.C. 
V5A 1S6, Canada}
\maketitle

\begin{abstract}
We study tunneling through an edge state formed around an antidot
in the fractional quantum Hall regime using 
Wen's chiral Luttinger liquid
theory extended to include mesoscopic effects. 
We identify a new regime 
where the Aharonov-Bohm oscillation amplitude 
exhibits a distinctive crossover 
from Luttinger liquid power-law
behavior to Fermi-liquid-like behavior as the
temperature is increased. 
Near the crossover temperature the amplitude has a pronounced
maximum. This non-monotonic behavior and  novel high-temperature 
nonlinear phenomena that we also predict provide new ways
to distinguish experimentally between Luttinger 
and Fermi liquids.
\end{abstract}


One of the most important outstanding questions in the study
of the quantum Hall
effect \cite{Prange and Girvin}
concerns the nature of the 
transport in the fractional regime. It has been established that 
for integral Landau-level filling factors, many aspects of the quantum
Hall effect can be understood in terms of Halperin's edge
states of the two-dimensional noninteracting electron gas 
\cite{Halperin}, 
and a useful
description of this is provided by the B\"uttiker-Landauer
formalism \cite{BL}. 
However, as was shown by Laughlin \cite{Laughlin}, 
the fractional quantum
Hall effect (FQHE) occurs because strong electron-electron
interactions lead to the formation of highly correlated
incompressible states at certain fractional filling factors.
In a large class of one-dimensional systems, 
interactions lead to a breakdown of
Fermi liquid theory and to the formation of a 
Luttinger liquid with a vanishing quasiparticle weight and
with, instead, 
bosonic low-energy 
excitations\cite{reviews,Haldane}. 
Transport in a Luttinger liquid was studied by
Kane and Fisher \cite{KF LL}, 
who have shown that the conductance of a
weakly disordered Luttinger liquid vanishes in the zero-temperature
limit, in striking contrast to a Fermi liquid.
The important connection between Luttinger liquids and the FQHE was made by
Wen \cite{Wen}, 
who used the Chern-Simons effective
field theory of the bulk FQHE \cite{Chern-Simons theory} 
to show that edge states in
the fractional regime should be {\it chiral} Luttinger liquids.
Wen's proposal has stimulated a considerable theoretical effort
to understand the properties of this exotic non-Fermi-liquid
state of matter 
\cite{Wen,Chamon and Wen,Moon etal,Pokrovsky and Pryadko,KF CLL,Fendley 
etal,Palacios and MacDonald}.

The first experimental evidence for a chiral Luttinger liquid
(CLL) was reported by 
Milliken {\it et al.} \cite{Milliken etal} on the tunneling
between FQHE edge states in a quantum point contact geometry.
As the gate voltage was varied, resonance peaks in the
conductance (caused by conditions of destructive
interference that prevent impurity-assisted tunneling between the two edge
channels) were observed that have the correct CLL temperature
dependence as predicted by Moon {\it et al.} \cite{Moon etal} 
and also by Fendley
{\it et al.} \cite{Fendley etal}.
In addition,
Chang {\it et al.} \cite{Chang etal}, working with a different 
type of system, have very recently reported experimental
evidence that is also in favor of CLL theory.
However, recent experiments by Franklin {\it et al.}
\cite{Franklin etal} on 
Aharonov-Bohm oscillations 
and by Maasilta and Goldman \cite{Maasilta and Goldman}
on resonant
tunneling in constrictions 
containing a quantum antidot are consistent
with Fermi liquid theory. This agreement with Fermi liquid theory
does not in itself rule out CLL theory because no detailed CLL
theory for the antidot geometry has been available. This has
motivated us to provide such a theory for the experimentally
realizable and analytically solvable strong-antidot-coupling
regime.

In this Letter, we study the Aharonov-Bohm (AB) effect in
the FQHE regime using CLL theory and extending it to include
mesoscopic effects. The problem we address may be realized
experimentally by measuring the tunneling through an edge
state formed around a quantum antidot, as in the experiments of
Franklin {\it et al.} \cite{Franklin etal} and of 
Maasilta and Goldman \cite{Maasilta and Goldman}.
(However, as explained below, our CLL theory is applicable
in a regime different from the one in which these
experiments were carried out.)
We begin by briefly summarizing our results:
The finite size of the antidot introduces a new temperature
scale, 
\begin{equation}
T_0 \equiv { \hbar v \over \pi k_{\rm B} L},
\label{T0}
\end{equation}
where $v$ is the edge-state Fermi velocity and $L$ is the circumference 
of the antidot edge state.
For example, a Fermi velocity of 
$10^6 \ {\rm cm/s}$ and circumference of
$1 \ \mu {\rm m}$ yields an experimentally 
accessible $T_0 \approx 20 {\rm mK}.$
In the strong-antidot-coupling regime,
CLL theory for filling factor $1/q \ (q \ {\rm odd})$ predicts the
low-temperature $(T \ll T_0)$ AB oscillation amplitude to
vanish with temperature as $T^{2q-2}$, in striking contrast
with Fermi liquid theory $(q=1)$. For $T$ near $T_0$,
there is a pronounced maximum in the AB amplitude, also in contrast
to a Fermi liquid. At high temperatures $(T \gg T_0)$, however,
we predict a new crossover to a $T^{2q-1} e^{-q T/T_0}$ temperature
dependence, which is qualitatively similar to Fermi
liquid behavior.
Experiments in the strong-antidot-coupling regime should be able
to distinguish between a Fermi liquid and our predicted nearly
Fermi liquid scaling. The finite size of the antidot also leads
to a remarkable {\it high-temperature} nonlinear response regime,
where the voltage $V$ satisfies $V \gg T \gg T_0$,
which may also be used to distinguish between Fermi liquid and
CLL behavior.

To study mesoscopic effects associated with edge states in the FQHE,
we shall extend CLL theory to include finite-size
effects. Finite-size effects in nonchiral Luttinger liquids have been
addressed in Refs.~\cite{Haldane} and \cite{Loss}. 
To proceed
in the chiral case we bosonize the electron field operators 
$\psi_{\pm}$ according to the convention
\begin{equation}
\rho_\pm = \pm { \partial_x \phi_{\pm} \over 2 \pi},
\label{bosonization convention}
\end{equation}
where 
$\rho_{\pm}$
is the normal-ordered charge density and $\phi_\pm$ is a chiral scalar
field for right (+) or left (--) movers. The dynamics of $\phi_\pm$ 
is governed by Wen's Euclidian action \cite{Wen}
\begin{equation}
S_\pm = {1 \over 4 \pi g} \int_0^L \! \! dx \int_0^\beta \! \! d\tau 
\ \partial_x \phi_\pm \big( \pm i \partial_\tau \phi_\pm +
v \partial_x \phi_\pm \big),
\label{euclidian action}
\end{equation}
where $g=1/q \ (q \ {\rm odd})$ is the bulk filling factor and $v$ is the
Fermi velocity. 
Here $L$ is the size (i.e., length) of a given edge state.
The field theory described by (\ref{euclidian action})
may be canonically quantized by imposing the equal-time commutation 
relation (modulo periodic extension)
\begin{equation}
[ \phi_{\pm}(x) , \phi_{\pm}(x')] = \pm i \pi g \ \! {\rm sgn}(x-x').
\label{full commutation relation}
\end{equation}
We then decompose $\phi_{\pm}$ into a nonzero-mode contribution 
${\bar \phi_\pm}$ satisfying periodic boundary conditions that 
describes the neutral excitations, and a zero-mode contribution 
$\phi_{\pm}^0$ that describes charged excitations: 
$\phi_{\pm} = {\bar \phi_\pm} + \phi_{\pm}^0$.
The nonzero-mode contribution may be expanded in a basis of Bose
annihilation and creation operators in the usual fashion,
\begin{equation}
{\bar \phi_{\pm}}(x) = \sum_{k \neq 0} \theta(\pm k) 
\sqrt{2 \pi g \over |k| L} 
\big( a_k e^{ikx} + a_k^\dagger e^{-ikx} \big) e^{-|k|a/2} ,
\label{nonzero-mode expansion}
\end{equation}
with coefficients determined by the requirement that 
${\bar \phi_{\pm}}$ itself satisfies 
(\ref{full commutation relation}) in the 
$L \rightarrow \infty$ limit.
In a finite-size system, however,
\begin{equation}
[{\bar \phi_\pm}(x),{\bar \phi_\pm}(x')] 
= \pm i \pi g \ \! {\rm sgn}(x-x') \mp {2 \pi i g \over L} (x-x'),
\label{nonzero-mode commutation relation}
\end{equation}
so we must require the zero-mode contribution to satisfy
\begin{equation}
[\phi_{\pm}^0(x), \phi_{\pm}^0(x')] = \pm {2 \pi i g \over L} (x-x')
\label{zero-mode commutation relation}
\end{equation}
for the total field to satisfy (\ref{full commutation relation}).
An expansion analogous to (\ref{nonzero-mode expansion}) for
$\phi_\pm^0$ may be constructed from the condition
(\ref{zero-mode commutation relation}) and, in addition, the requirement
\begin{equation}
\phi_{\pm}^0(x+L) - \phi_\pm^0(x) = \pm 2 \pi N_\pm ,
\label{zero-mode boundary condition}
\end{equation}
which follows from (\ref{bosonization convention}), where 
$N_\pm \equiv \int_0^L dx \ \rho_{\pm}$ is the charge of an excited
state relative to the ground state. 
Conditions (\ref{zero-mode commutation relation}) and 
(\ref{zero-mode boundary condition}) together determine $\phi_\pm^0$,
up to an additive c-number constant, as
\begin{equation}
\phi_\pm^0(x) = \pm {2 \pi \over L} N_\pm x - g \ \! \chi_\pm ,
\label{zero-mode expansion}
\end{equation}
where $\chi_\pm$ is a phase operator conjugate to 
$N_\pm$ satisfying $[\chi_\pm, N_\pm] = i.$
Equations (\ref{nonzero-mode expansion}) and (\ref{zero-mode expansion})
may now be used to write the normal-ordered Hamiltonian corresponding
to (\ref{euclidian action}) as
\begin{equation}
H_\pm = {v \over 4 \pi g} \int_0^L dx \big(\partial_x \phi_\pm \big)^2
= {\pi v \over g L} N_\pm^2 + \sum_{k \neq 0} \theta(\pm k) v |k| 
a_k^\dagger a_k .
\label{hamiltonian}
\end{equation}
In a finite-size system, the level spacing for neutral and charged
excitations scale with system size as $1/L$, and they become gapless in 
the $L \rightarrow \infty$ limit.

What are the allowed eigenvalues of $N_\pm$? The answer may be determined
by
bosonization: To create an electron, we need a $\pm 2 \pi$ kink in
$\phi_\pm$. The electron field operators can therefore be bosonized as 
\begin{equation}
\psi_\pm(x) = {1\over \sqrt{2 \pi a}} \,
e^{i [\phi_\pm(x)  \pm {\pi x \over L }]/g},
\label{bosonization}
\end{equation}
where $a$ is the same microscopic cutoff length that appears
in (\ref{nonzero-mode expansion}). The additional c-number 
phase factor is chosen for convenience.
To see that (\ref{bosonization}) is valid, note that
$[\rho_\pm(x) , \psi_\pm^\dagger(x')] 
= \delta(x-x') \psi_\pm^\dagger(x')$, so $\psi_\pm^\dagger(x)$ creates
an electron at position $x$. 
Equation (\ref{bosonization}) implies that $\psi_\pm(x+L)=\psi_\pm(x) 
e^{\pm i2\pi N_{\pm}/g}$. Thus, periodic boundary conditions
on $\psi_\pm(x)$ lead to the important result that the allowed
eigenvalues of $N_{\pm}$ are given by
\begin{equation}
N_\pm = n g ,
\label{fractional charge}
\end{equation}
which means that there exists {\it fractionally charged} excitations or
quasiparticles,
as expected in a FQHE system.

Coupling to an AB flux $\Phi$ is achieved by adding
$\delta {\cal L} = {1 \over c} j_\pm A$ to the Lagrangian, where 
$j_\pm = \pm {e \over 2 \pi} \partial_t \phi_\pm$ is the
bosonized current density and $A$ is a vector potential. The flux
couples only to the zero modes, and results in the replacement
$N_\pm^2 \rightarrow (N_\pm \pm g \Phi/\Phi_0)^2$ 
in (\ref{hamiltonian}),
where $\Phi_0 \equiv hc/e$.
The grand-canonical partition function of the mesoscopic edge state 
factorizes into a zero-mode contribution,
$ Z^0 = \sum_n e^{- \pi g v  
(n-\Phi/\Phi_0)^2 / LT }, $
which depends on $\Phi$, and a 
flux-independent contribution
from the nonzero modes \cite{persistent currents}.
Note that if the $N_\pm$ were restricted to be integral,
then the partition function and the associated grand-canonical
potential would be periodic functions of
flux with period 
$\Phi_0/g$.
The fractionally charged excitations
(\ref{fractional charge}) are therefore responsible for
restoring the AB period to the proper value $\Phi_0$, as is known in
other contexts \cite{AB period}. 

We begin our study of transport by performing a perturbative
renormalization group (RG) analysis in the weak-antidot-coupling
regime (see Fig.~1a). 
In this case, $S = S_0 + \delta S$, where
$S_0 \equiv S_{\rm L} + S_{\rm R} 
+ S_{\rm A}$ is the sum of actions of the form
(\ref{euclidian action}) for the left moving, right moving,
and antidot edge states, respectively, and
$\delta S \equiv \sum_m \int_\tau (V_{+} + V_{-} + {\rm c.c.} )$
is the weak coupling between them. 
Here $V_\pm(\tau) \equiv \Gamma_\pm^{(m)} 
e^{i m \phi_\pm(x_\pm,\tau)} e^{-im \phi_A(x_\pm,\tau)} / 2 \pi a $
describes the tunneling of
$m$ quasiparticles from an incident edge state into the
antidot edge state at point $x_\pm$ with dimensionless amplitude
$\Gamma_\pm^{(m)}$ \cite{KF LL}.
We assume the leads, described by $S_{\rm L}$ and
$S_{\rm R}$, to be macroscopic, and we also assume for simplicity that
$|\Gamma_-^{(m)}|=|\Gamma_+^{(m)}|$.
We shall need the correlation function
$C_\pm(x,\tau) \equiv
\langle e^{i m \phi_\pm(x,\tau)} e^{-im \phi_\pm(0)} \rangle$
taken with respect to $S_0$,
which, at zero temperature and for values of $x$ such that
$x \ll L$, is given by
\begin{equation}
C_\pm(x,\tau) = \bigg( { \pm ia \over  x \pm iv \tau \pm ia} 
\bigg)^{2 \Delta} ,
\label{correlation function}
\end{equation}
where $\Delta = m^2 g / 2$ is the scaling dimension of
$e^{i m \phi_{\pm}}$.

Consider now the correlation function
\begin{equation}
\big\langle V_+^\dagger(\tau) V_+(0) \big\rangle
= {  \big| \Gamma_+^{(m)} \big|^2 \over 4 \pi^2 a^2}
\big\langle e^{-i m \phi_{+}(x_+,\tau)} 
e^{im \phi_{+}(x_+,0)} \big\rangle
\big\langle e^{i m \phi_{A}(x_+,\tau)} 
e^{-im \phi_{A}(x_+,0)} \big\rangle,
\label{VV correlation function}
\end{equation}
which arises in a perturbative calculation of the
total partition
function
$Z = \int {\cal D}\phi_L {\cal D}\phi_R {\cal D}\phi_A e^{-S}$.
For $Z$ to be invariant under a small decrease in cutoff 
$a \rightarrow a' = s a$, we need 
$\Gamma' = s^{1-2 \Delta} \Gamma$, or
\begin{equation}
{d \Gamma_{+}^{(m)} \over d \ell} = \big( 1 - m^2 g \big) 
\Gamma_{+}^{(m)},
\label{flow equations}
\end{equation}
where $\ell \equiv \ln (a/a')$.
$\Gamma_{-}^{(m)}$ satisfies an identical RG equation.
These leading-order flow equations, 
which show that quasiparticle $(m=1)$
backscattering processes are relevant and electron $(m=1/g)$
backscattering is irrelevant when $g=1/3$, were first derived
by Kane and Fisher \cite{KF LL} using momentum-shell RG.

Next consider the correlation function
$ \langle V_+^\dagger(\tau) V_+(0) 
V_-^\dagger(\tau') V_-(0) \rangle$, which arises in fourth order.
A Wick expansion gives {\it local} terms as in 
(\ref{VV correlation function}), and, in addition, {\it nonlocal}
antidot correlation functions like
$\langle e^{im \phi_{A}(x,\tau)} e^{-im \phi_{A}(0)} \rangle$
with $x \neq 0$. 
Equation (\ref{correlation function}) shows that
the nonlocal terms (for $x \ll L$, with $L$ now the
size of the antidot edge state) scale in the same way as the local
terms \cite{RG footnote}. 
The Kane-Fisher flow equations (\ref{flow equations}) are
therefore valid in the antidot problem considered here.

This scaling analysis shows that 
off resonance \cite{resonance footnote} and
at low temperatures the antidot will be in the strongly coupled regime
shown in Fig.~1b. Furthermore, if the antidot system starts in the
strongly coupled regime (by an appropriate choice of gate voltages), 
it will stay in this regime 
because the $m=1$ quasiparticle backscattering process
(which would be relevant in the RG sense) is not allowed
in this edge-state configuration and only electrons can tunnel.
The strong-antidot-coupling regime therefore admits a perturbative
treatment \cite{Wen pert theory}, to which we now turn.
Details of the calculations shall be given elsewhere.

The current 
passing between edge states $L'$ and $R'$, driven by
their potential difference $V$, is defined by (restoring units)
$I \equiv -e \langle {\dot N}_{L'} (t)\rangle$, where
$N_{L'}$ is the charge of edge state $L'$ as defined before
(\ref{zero-mode expansion}). The current is now evaluated for 
small tunneling amplitudes $\Gamma_i \ (i=1,2)$, which for simplicity 
are taken to
be equal apart from AB phase factors \cite{voltage footnote}. 
The result is
\begin{equation}
I = - 2 |\Gamma|^2 \ \! {\rm Im} \ \! \bigg[ 
X_{11}(\omega) + X_{22}(\omega) 
+ e^{i2 \pi \Phi / \Phi_0} \ \! X_{12}(\omega) 
+ e^{-i2 \pi \Phi / \Phi_0} \ \! X_{21}(\omega) 
\bigg]_{\omega = V},
\label{IV relation}
\end{equation} 
where $X_{ij}(\omega)$ is the Fourier transform of 
$X_{ij}(t) \equiv -i \theta(t) \langle [B_i(t), B_j^\dagger(0)] \rangle$
and
$B_i \equiv \psi_L(x_i) \psi_R^\dagger(x_i)$
is an electron tunneling operator acting at point $x_i$.
This response function can be calculated using bosonization techniques
and the result for filling factor $1/q$ is
\begin{equation}
X_{ij}(t) = - \theta(t) { a^{2q-2} \over 2 \pi^2} \ {\rm Im} \ \!
{ (\pi / L_{\rm T} )^{2q} 
\over \sinh^q [\pi (x_i-x_j+vt+ia)/L_{\rm T}]
\sinh^q [\pi (x_i-x_j-vt-ia)/L_{\rm T}] } \,\ ,
\label{response function}
\end{equation}
where 
$L_{\rm T} \equiv \beta v$ is the thermal length.
Each term $X_{ij}$ in (\ref{IV relation}) corresponds to a process
occurring 
with a probability proportional to $|\Gamma_i \Gamma_j|$. 
The {\it local} terms $X_{11}$ and $X_{22}$ therefore
describe {\it independent} tunneling at $x_1$ and $x_2$,
respectively, whereas the
{\it nonlocal} terms $X_{12}$ and $X_{21}$ describe {\it coherent} 
tunneling through both points. 
The AB phase naturally couples only to the latter.
We shall see that the local contributions behave exactly like the
tunneling in a quantum point contact.
The AB effect, however, is a consequence of the nonlocal terms, and
we shall show that there are new  
non-Fermi-liquid phenomena associated with these terms
that are directly accessible to experiment.

We have Fourier transformed (\ref{response function}) exactly and find a
crossover behavior in the nonlocal response functions
when the thermal length $L_T$ becomes less than 
$|x_i-x_j|$.
The finite size of the antidot therefore provides 
an important new temperature scale given in Eqn.~(\ref{T0}). 
Note that
$T_0$ is closely related to the energy level 
spacing $\Delta \epsilon 
\equiv 2 \pi v /L$ for noninteracting electrons with linear dispersion
in a ring of circumference $L$: $T_0=\Delta \epsilon /2\pi^2$.
The current in the strong-antidot-coupling regime
can generally be written as
$ I = I_0 + I_{\rm AB} \cos(2 \pi \Phi/\Phi_0),$
where $I_0$ is the {\it direct} contribution resulting from the local
terms and $I_{\rm AB}$ is the AB contribution resulting from the
nonlocal terms. 
For noninteracting electrons, the B\"uttiker-Landauer  
formula or our perturbation theory with
$q=1$ may be used to show that
$ I^{FL}_0 = |\Gamma |^2 V / \pi$
and
$ I^{FL}_{\rm AB} = 2 |\Gamma |^2 T \sinh^{-1}(T/T_0) \sin(VL/2v).$
The corresponding conductances are
$G^{\rm FL}_0 = |\Gamma |^2 / \pi$ and
$ G^{\rm FL}_{\rm AB} = (|\Gamma |^2 / \pi) (T/T_0) \sinh^{-1}(T/T_0).$

The exact current-voltage relation for the $q=3$ CLL is 
\begin{equation}
 I_0= {|\Gamma|^2 a^4 V \over 120 \pi v^6}
\bigg( 112 {\pi}^4 T^4+ 
40 {\pi}^2 T^2 V^2 + V^4 \bigg),
\end{equation}
and
\begin{eqnarray}
I_{\rm AB}= - {|\Gamma|^2 a^4 \pi^2 \over v^6} 
{ T^3 \over{\sinh^3(T/T_0)}}
\bigg\lbrace 
&\bigg[& V^2+ 4\pi^2 T^2 \bigg(1-3
\coth^2(T/T_0) \bigg) \bigg]
\sin\bigg({VL \over 2v}\bigg) 
\nonumber\\
&+& 6 \pi V T \coth(T /T_0) 
\cos\bigg({VL \over 2v}\bigg) 
\bigg\rbrace .
\label{IAB}
\end{eqnarray}
In the limit $L \rightarrow 0$, $I_{\rm AB}$ always reduces to
$I_0$.
The AB conductance is 
\begin{equation}
G_{\rm AB} = -{2 \pi^3 | \Gamma |^2 a^4 \over v^6} 
{ T^4 \over \sinh^3(T/T_0)}
\bigg\lbrace 3 \coth \bigg({T \over T_0}\bigg) 
+ \bigg({T \over T_0}\bigg) \bigg[
1- 3 \coth^2 \bigg({T \over T_0}\bigg) \bigg] \bigg\rbrace,
\end{equation}
which is shown in Fig.~1c along with the
corresponding Fermi-liquid result..

We now summarize our results for general $q$.
The complete phase diagram is very rich and will be described
in detail elsewhere.
Here we shall summarize the transport properties
as a function of temperature for fixed voltage, first for 
$V \ll T_0$ and then for $V \gg T_0$. 

{\sl Low-voltage} $(V \ll T_0)$ {\sl regime}:
There are three temperature regimes here. When
$T \ll V \ll T_0$, both $I_0$ and $I_{\rm AB}$ have nonlinear
behavior, varying with voltage as $V^{2q-1}$. 
When the temperature
exceeds $V$, the response becomes linear.
When $V \ll T \ll T_0$, both $G_0$ and $G_{\rm AB}$ vary with
temperature as 
\begin{equation}
G \propto \bigg({T \over T_F}\bigg)^{2q-2},
\label{low-temperature G}
\end{equation}
in striking contrast to a Fermi liquid $(q=1)$.
This is the same low-temperature power-law scaling 
predicted \cite{KF LL,Moon etal,Fendley etal} and 
observed \cite{Milliken etal} 
in a quantum point contact tunneling geometry.
Here $T_{\rm F} \equiv v/a$ is an effective Fermi temperature.
Near $T \approx 2T_0$ for the $q=3$ case, we find 
that $G_{\rm AB}$ displays a 
pronounced maximum, also in contrast to a 
Fermi liquid (see Fig.~1c).
Increasing the temperature further, however, we cross over into the
$V \ll T_0 \ll T$ regime, where $G_0$ scales as in 
(\ref{low-temperature G}), but now 
\begin{equation}
G_{\rm AB} \propto \bigg({T \over T_0} \bigg)
\bigg({T \over T_{\rm F}} \bigg)^{2q-2} e^{-qT/T_0}.
\label{high-temperature G}
\end{equation}
Thus, the AB oscillation amplitude exhibits 
a crossover from the well-known
$T^{2q-2}$ Luttinger liquid behavior to a new scaling behavior
that is much closer to a Fermi liquid. 
However, as compared to the Fermi liquid case, the crossover
temperature here is effectively {\it lower} by a factor of $q$.
Careful measurements in this experimentally accessible regime should
be able to distinguish between a Fermi liquid and our 
predicted nearly Fermi-liquid temperature dependence.

{\sl High-voltage} $(V \gg T_0)$ {\sl regime}:
Again there are three temperature regimes. At the lowest
temperatures, $T \ll T_0 \ll V$, the response is 
nonlinear. The direct contribution varies with voltage as
$I_0 \propto V^{2q-1}$.
The AB current is more
complicated, involving power-laws times trigonometric 
functions of the ratio $V/T_0$. 
For the case $q=3$,
\begin{equation}
I_{\rm AB} = -{8 |\Gamma |^2 a^4 \over \pi v L^5}
\bigg\lbrace \bigg[{3V\over 2\pi T_0}\bigg]
\cos \bigg({V \over 2\pi T_0}\bigg)
 - \bigg[ 3 - \bigg({V\over 2\pi T_0}\bigg)^2 \bigg]
\sin \bigg({V \over 2\pi T_0} \bigg) \bigg\rbrace . 
\end{equation}
As the temperature is increased further to
$T_0 \ll T \ll V$, we find a crossover
to a remarkable {\it high-temperature} nonlinear regime. 
Here, $I_0 \propto V^{2q-1}$ as before, but now
\begin{equation}
I_{\rm AB} \propto \bigg({T \over T_0}\bigg)^q e^{-qT/T_0} V^{q-1}
\sin \bigg({V \over 2 \pi T_0} \bigg).
\label{high-temperature nonlinear IAB}
\end{equation}
Note the additional $V^{q-1}$ term that is not present in the
corresponding Fermi liquid result.
Therefore, the nonlinear response
can also be used to distinguish between Fermi liquid and
CLL behavior, even at relatively high temperatures.
When the temperature exceeds V, the response finally
becomes linear. 
When $T_0 \ll V \ll T$, $G_0$ scales as in 
(\ref{low-temperature G}), whereas $G_{\rm AB}$ scales
as in (\ref{high-temperature G}). 
Thus, at high temperatures the low- and high-voltage 
regimes behave similarily.

In conclusion, we have studied the AB effect for filling factor
$1/q \ (q \ {\rm odd})$ in the strong-antidot-coupling limit
with CLL theory.
The low-temperature linear response is similar to that in a 
quantum point contact. 
However, the AB oscillations are a mesoscopic effect 
and, as such, are diminished in amplitude above a crossover temperature
$T_0$ determined by the size of the antidot. Above $T_0$, the
temperature dependence of the AB oscillations is qualitatively
similar to that in a Fermi liquid (see Fig.~1c). It is clear
that a related crossover occurs in the weak-antidot-coupling
regime as well. In addition, we have identified a new high-temperature
nonlinear response regime that may also be used to distinguish
between a Fermi and Luttinger liquid.
 
We thank Leonid Pryadko for useful discussions.
This work has been supported by NSERC of Canada.


\begin{figure}
\caption{(a) Aharonov-Bohm effect geometry in the  
weak-antidot-coupling regime. The solid lines represent edge
states and the dashed lines
denote weak tunneling points.
(b) Edge-state configuration in the strong-antidot-coupling
regime. Here the edge states are almost completely reflected.
(c) Temperature dependence of $G_{\rm AB}$ for
the cases $q=1$ (dashed curve) and $q=3$ (solid curve). 
Each curve is normalized to have unit amplitude at its
maximum.}
\label{figure1}
\end{figure}


\begin{references}
\bibitem{persistent currents} The chiral \lq\lq persistent
currents'' associated with FQHE edge states have
a universal non-Fermi-liquid temperature dependence and
will be discussed elsewhere.
\bibitem{RG footnote} Nonlocal terms with $x \approx L$ are always
irrelevant.
\end{references}
\end{document}